\documentclass{cupconf}
\usepackage[hyperindex,breaklinks]{hyperref}
\usepackage{graphicx}
\usepackage{graphicx}
\usepackage{epsf}

  \checkfont{eurm10}
  \iffontfound
    \IfFileExists{upmath.sty}
      {\typeout{^^JFound AMS Euler Roman fonts on the system,
                   using the 'upmath' package.^^J}%
       \usepackage{upmath}}
      {\typeout{^^JFound AMS Euler Roman fonts on the system, but you
                   dont seem to have the}%
       \typeout{'upmath' package installed. cupconf.cls can take advantage
                 of these fonts,^^Jif you use 'upmath' package.^^J}%
      }
  \else
  \fi

  \checkfont{msam10}
  \iffontfound
    \IfFileExists{amssymb.sty}
      {\typeout{^^JFound AMS Symbol fonts on the system, using the
                'amssymb' package.^^J}%
       \usepackage{amssymb}%
       \let\le=\leqslant  \let\leq=\leqslant
         \let\geq=\geqslant
      }{}
  \fi

  \IfFileExists{amsbsy.sty}
    {\typeout{^^JFound the 'amsbsy' package on the system, using it.^^J}%
     \usepackage{amsbsy}}
    {\providecommand\boldsymbol[1]{\mbox{\boldmath $##1$}}}

\newsavebox{\astrutbox}
\sbox{\astrutbox}{\rule[-5pt]{0pt}{20pt}}

\newcommand\etal{\mbox{\textit{et al.}}}

\title[Simulations in general relativity]
{Black hole formation and growth:
simulations in general relativity}

\author[S. L. Shapiro]%
{S\ls T\ls U\ls A\ls R\ls T\ns L.\ns S\ls H\ls A\ls P\ls I\ls R\ls O$^{1,2}$}

\affiliation{$^1$Department of Physics, University of Illinois at
Urbana-Champaign, Urbana, IL 61801, USA\\[\affilskip]
$^2$Department of Astronomy and National Center for Supercomputing
Applications, University of Illinois at
Urbana-Champaign, Urbana, IL 61801, USA}

\pubyear{2008}
\volume{?}
\pagerange{1-15}
\date{?? and in revised form ??}
\begin{document}

\maketitle

\begin{abstract}
Black holes are popping up all over the place: in compact 
binary X-ray sources and GRBs, in quasars, AGNs and the cores
of all bulge galaxies, in binary black holes and binary black hole-neutron
stars, and maybe even in the LHC! Black holes are strong-field objects
governed by Einstein's equations of general relativity. Hence
general relativistic, numerical simulations of dynamical phenomena
involving black holes may help reveal ways in which black holes can 
form, grow and be detected in the universe. To convey the state-of-the art, 
we summarize several representative simulations here, including the 
collapse of a hypermassive neutron star to
a black hole following the merger of a binary neutron star, 
the magnetorotational collapse of a massive star to a black hole, 
and the formation and growth of supermassive black hole
seeds by relativistic MHD accretion in the early universe.
\end{abstract}

\firstsection 
\section{Introduction}

Black holes are `sighted' everywhere in the universe these days.
Originally located in compact binary X-ray sources in the 1970's, 
the cosmic presence of black holes has expanded considerably in recent decades. 
They now are believed to be the central engines that power quasars, 
active galactic nuclei (AGNs) and gamma-ray bursts (GRBs). They
are identified in the cores of all bulge galaxies. They 
are presumed to form signficant populations of compact binaries, including 
black hole-black hole binaries (BHBHs) and black hole-neutron star 
binaries (BHNSs).  Black holes may even show up soon in the 
Large Hadron Collider!

Gravitationally, black holes are strong-field objects whose properties
are governed by Einstein's theory of relativistic gravitation --- general
relativity.  General relativistic simulations of gravitational collapse to
black holes, BHBH mergers and recoil, black hole accretion, and
other astrophysical phenomena involving black holes may help
reveal how, when and where black holes form, grow and interact in the 
physical universe.  As a consequence, such simulations can help identify
the ways in which black holes can best be detected. 


To illustrate how our understanding of black hole phenomena is sharpend by
large-scale simulations in general relativity, 
we summarize in this paper the results of several recent computational
investigations. The first few involve the formation of black holes from
stellar collapse, while the last one concerns supermassive black hole growth
via disk accretion. Most of these simulations 
utilize the tools of numerical 
relativity to solve Einstein's field equations of general relativity.
So we shall begin 
with a brief overview of this important, rapidly maturing field.

\section{Numerical Relativity and the $3+1$ Formalism}

\begin{figure}
       \centerline{\includegraphics[height=3.3 cm]
                                   {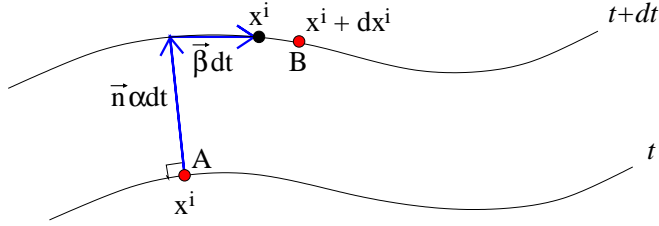}}
   \caption{3+1 decomposition of spacetime.}
   \label{ADM}
   \end{figure}

Most current work in numerical relativity is performed within the framework of
the $3+1$ decomposition of Einstein's field equations using some
adaptation of the standard ADM equations 
(\cite{ArnDM62}). 
In this framework spacetime is
sliced up into a sequence of spacelike hypersurfaces of constant
time $t$, appropriate for solving an initial-value problem.
Consider two such time slices separated by an infinitesimal interval $dt$ as 
shown in Fig~\ref{ADM}.  
The spacetime metric measures the invariant interval between neighboring 
points $A$ and $B$ on the two slices according to
\begin{equation}
ds^2 = -\alpha^2 dt^2 +
\gamma_{ij}(dx^i+\beta^i dt) (dx^j+\beta^j dt)\ ,
\end{equation}
where $\gamma_{ij}$ is the spatial 3-metric 
on a time slice, $\alpha$ is the lapse function determining
the proper time between the slices as measured by a time-like normal observer 
$n^a$ at rest in the slice, and $\beta^i$ is the  
shift vector, a spatial 3-vector 
that describes the relabeling of the spatial coordinates of points in the slice.
The gravitational field satisfies the Hamiltonian and
 momentum {\it constraint equations} 
on each time slice, including the initial slice at $t=0$:
\begin{eqnarray}
R + K^2 - K_{ij}K^{ij} &=& 16 \pi \rho
~~({\rm Hamiltonian})\ , \label{ham} \\
D_j(K^{ij} - \gamma^{ij}K) &=& 8 \pi S^i
~~({\rm momentum})\ \label{mom}.
\end{eqnarray}
Here $R=R^i{}_i$ is the scalar curvature on the slice,
$R_{ij}$ is the 3-Ricci tensor, $K_{ij}$ is the 
extrinsic curvature, $K$ is its trace, and $D_j$ is the covariant
derivative operator on the slice. The quantities $\rho$ and $S^i$ are
the mass and momentum densities of the matter, respectively; 
such matter source terms
are formed by taking suitable projections of the matter stress-energy tensor
$T^{ab}$ with respect to the normal observer. Included in this stress-energy
tensor are contributions from all the nongravitational sources of mass-energy,
which we are simply calling the ``matter'' (e.g., baryons, 
electromagnetic fields, neutrinos, etc.).

A gravitational field  satisfying the constraint equations on the initial
slice can be determined at future times by integrating the 
{\it evolution equations},
\begin{eqnarray}
\partial_t \gamma_{ij} &=& -2 \alpha K_{ij} + D_i \beta_j
   + D_j \beta_i\ , \label{dgdt}\\
\partial_t K_{ij} &=& \alpha (R_{ij} - 2 K_{ik} K^k_{~j}
 + K K_{ij})
- D_i D_j \alpha \label{dKdt} \\
&+& \beta^k \partial_k K_{ij} + K_{ik} \partial_j \beta^k
 + K_{kj} \partial_i \beta^k 
 - 8 \pi \alpha (S_{ij} - \frac{1}{2} \gamma_{ij} (S - \rho))\ , \nonumber
\end{eqnarray}
where $S$ and $S_{ij}$ are additional matter source terms.
The evolution equations guarantee that the field equations will 
automatically  satisfy the constraints on all future time slices 
identically, provided they satisfy them on the 
initial slice. Of course, this statement applies to the analytic
set of equations and not necessarily to their numerical counterparts.

Note that the $3+1$ formalism prescribes no equations 
for $\alpha$ and $\beta^i$. These four functions embody the four-fold
gauge (coordinate) freedom inherent in general relativity. 
Choosing them judiciously, especially in the presence of black holes,
is one of the main challenges of numerical relativity.

\subsection{The BSSN Scheme}

During the past decade, significant improvement in our ability to 
numerically integrate Einstein's equations stably in full $3+1$ dimensions 
has been achieved by recasting the original ADM system of equations. 
One such reformulation is the so-called 
BSSN scheme (\cite{ShiN95}; \cite{BauS99}). In this scheme, the physical metric and extrinsic
curvature variables are replaced in favor of the conformal metric and 
extrinsic curvature, in the spirit of 
the ``York-Lichnerowicz'' split (\cite{Lic44}; \cite{Yor71}):
\begin{eqnarray}
\tilde{\gamma}_{ij} &=& e^{-4\phi} \gamma_{ij},
~~~{\rm where}~~~ e^{4\phi} = \gamma^{1/3}\ ,\\
\tilde{A}_{ij} &=& \tilde{K}_{ij}- \frac{1}{3}\tilde{\gamma}_{ij} K\ .
\end{eqnarray}
Here a tilde $\tilde{}$ denotes a conformal quantity and $\gamma$ is the 
determinant of $\gamma_{ij}$.
At the same time, a connection function $\tilde{\Gamma}^i$ is introduced
according to
\begin{eqnarray} \label{Gamma}
\tilde{\Gamma}^i \equiv \tilde{\gamma}^{jk}\tilde{\Gamma}^i{}_{jk}
   = - \partial_j \tilde{\gamma}^{ij}\ .
\end{eqnarray}
The quantities that are independently evolved in this scheme are now
$\tilde{\gamma}_{ij}, \tilde{A}_{ij}, \phi, K$ and $\tilde{\Gamma}^i$.
The advantage is that the Riemann operator appearing in 
the evolution equations (cf. eqn.~(\ref{dKdt})) takes 
on the form,
\begin{eqnarray}
\tilde{R}_{ij} =
-\frac{1}{2} \underbrace{\tilde{\gamma}^{lm}\partial_m \partial_l
\tilde{\gamma}_{ij}}_{`Laplacian`} +
\underbrace{\tilde{\gamma}_{k(i}\partial_{j)}\tilde{\Gamma}^k}_
{\rm remaining\ 2nd\ derivs} + \cdots\ .
\end{eqnarray}
Thus the principal part of this operator, 
$\tilde{\gamma}^{lm}\partial_m \partial_l \tilde{\gamma}_{ij}$ is that of a
Laplacian acting on the components of the 
metric $\tilde {\gamma}_{ij}$.
All the other second derivatives of the metric have been absorbed in the 
derivatives of the connection functions. The coupled evolution equations
for $\tilde{\gamma}_{ij}$ and $\tilde{A}_{ij}$
 (cf. eqns.~(\ref{dgdt}) and (\ref{dKdt}) then reduce essentially to a
wave equation,
\begin{eqnarray}
\partial_t^2 \tilde{\gamma}_{ij}
\sim \partial_t \tilde{A}_{ij}
\sim \tilde{R}_{ij} \sim  \nabla^2  \tilde{\gamma}_{ij}\ .
\end{eqnarray}
Wave equations not only reflect the hyperbolic nature of general relativity,
but can be implemented numerically in a straight-forward and stable manner.
By now, numerous simulations 
have demonstrated the
dramatically improved stability achieved in the BSSN scheme
over the standard ADM equations,
and considerable effort has gone into explaining the
improvement on theoretical grounds [see, e.g., \cite{BauS03} for discussion and
references]. Many of the recent BHBH merger
calculations have been performed using this scheme, beginning with 
\cite{CamLMZ06} and \cite{BakCCKV06} (but see \cite{Pre05} for an alternative approach). The same is true for the simulations described below.

\section{Binary Neutron Star Mergers and Hypermassive Stars}
\label{bns}

The protagonist of several different astrophysical 
scenarios probed by recent numerical simulations is
a { \it hypermassive} star, typically a hypermassive neutron star (HMNS). 
A hypermassive
star is an equilibrium fluid configuration that supports itself against 
gravitational collapse by {\it differential} rotation. Uniform rotation
can increase the maximum mass of a nonrotating, spherical
equilibrium star by at most $\sim 20\%$, but differential rotation
can achieve a much higher increase (\cite{BauSS00}; \cite{MorBS04}).
Dynamical simulations using the BSSN scheme
demonstrate (\cite{BauSS00}) that hypermassive stars can be 
constructed that are {\it dynamically} stable, provided 
the ratio of rotational kinetic to gravitational potential energy, $\beta$,
is not too large; for $\beta \gtrsim 0.24$ the configuration is subject
to a nonaxisymmetric dynamical bar 
instability (\cite{ShiBS00}; \cite{SaiSBS01}).
However, all hypermassive stars are {\it secularly} unstable to
the redistribution of angular angular momentum by viscosity, magnetic
braking, gravitational radiation, or any other agent that dissipates 
internal shear. Such a 
redistribution tends to drive a hypermassive
star to uniform rotation, which cannot
support the mass against collapse. Hence hypermassive stars are 
transient phenomena. Their formation
following, for example, a NSNS merger,
or core collapse in a massive, rotating star, may ultimately 
lead to a `delayed' collapse to a black hole on secular (dissipative) 
timescales.  Such a collapse will be accompanied inevitably 
by a delayed gravitional wave burst (\cite{BauSS00}; \cite{Sha00}).

\begin{figure}
   \begin{center}
   \includegraphics[height=5.0cm]{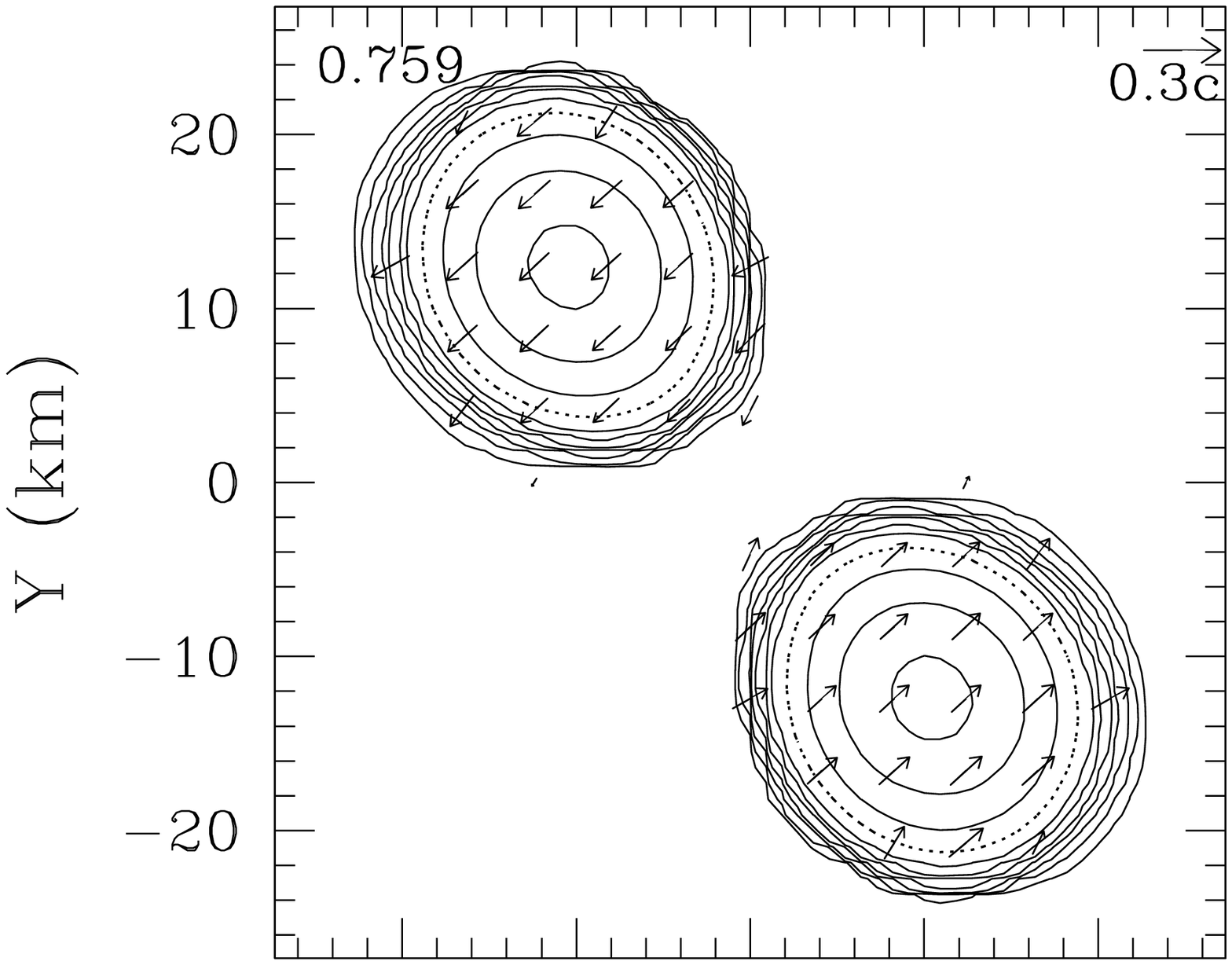}
   \includegraphics[height=5.0cm]{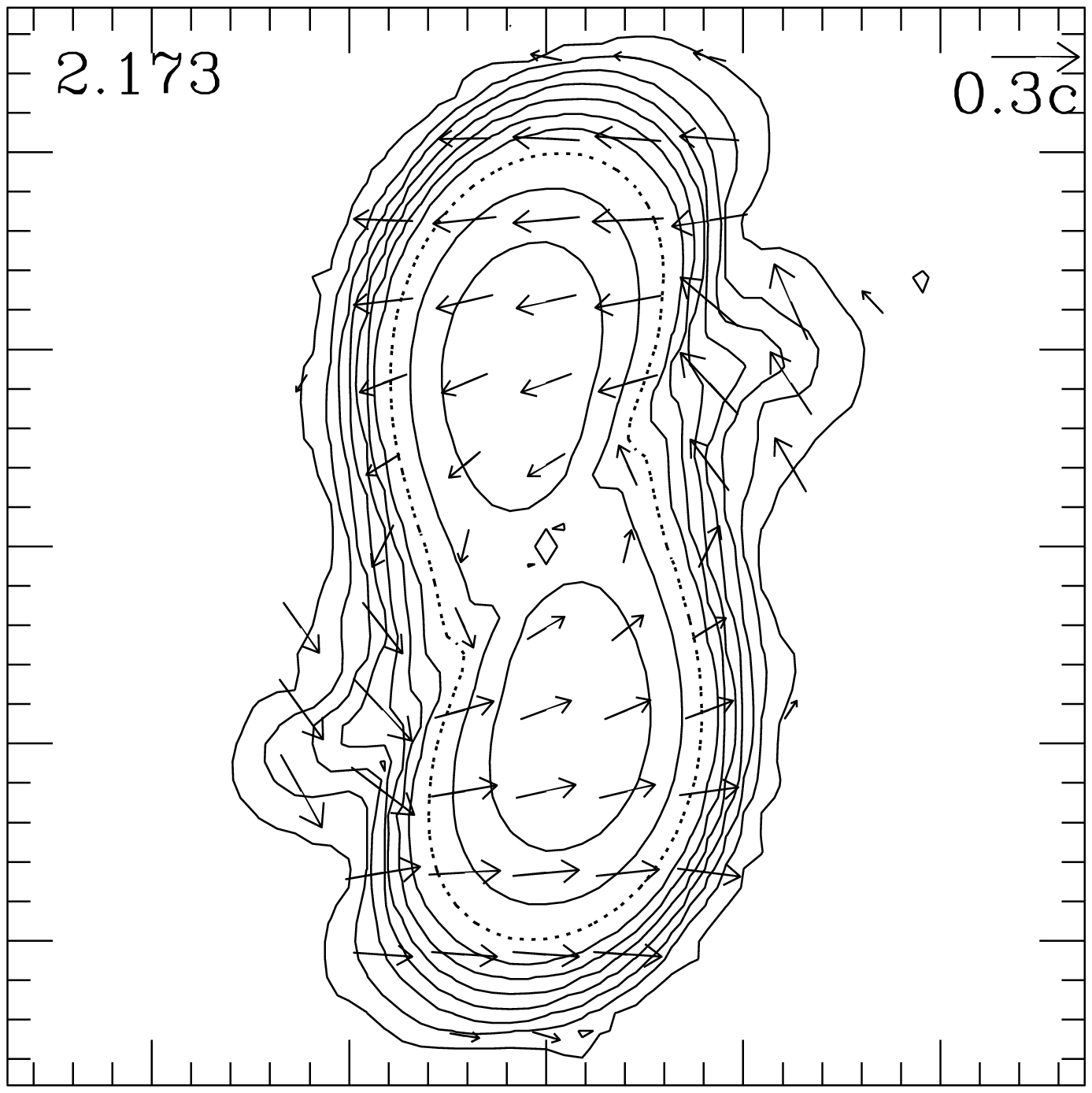}
   \includegraphics[height=5.0cm]{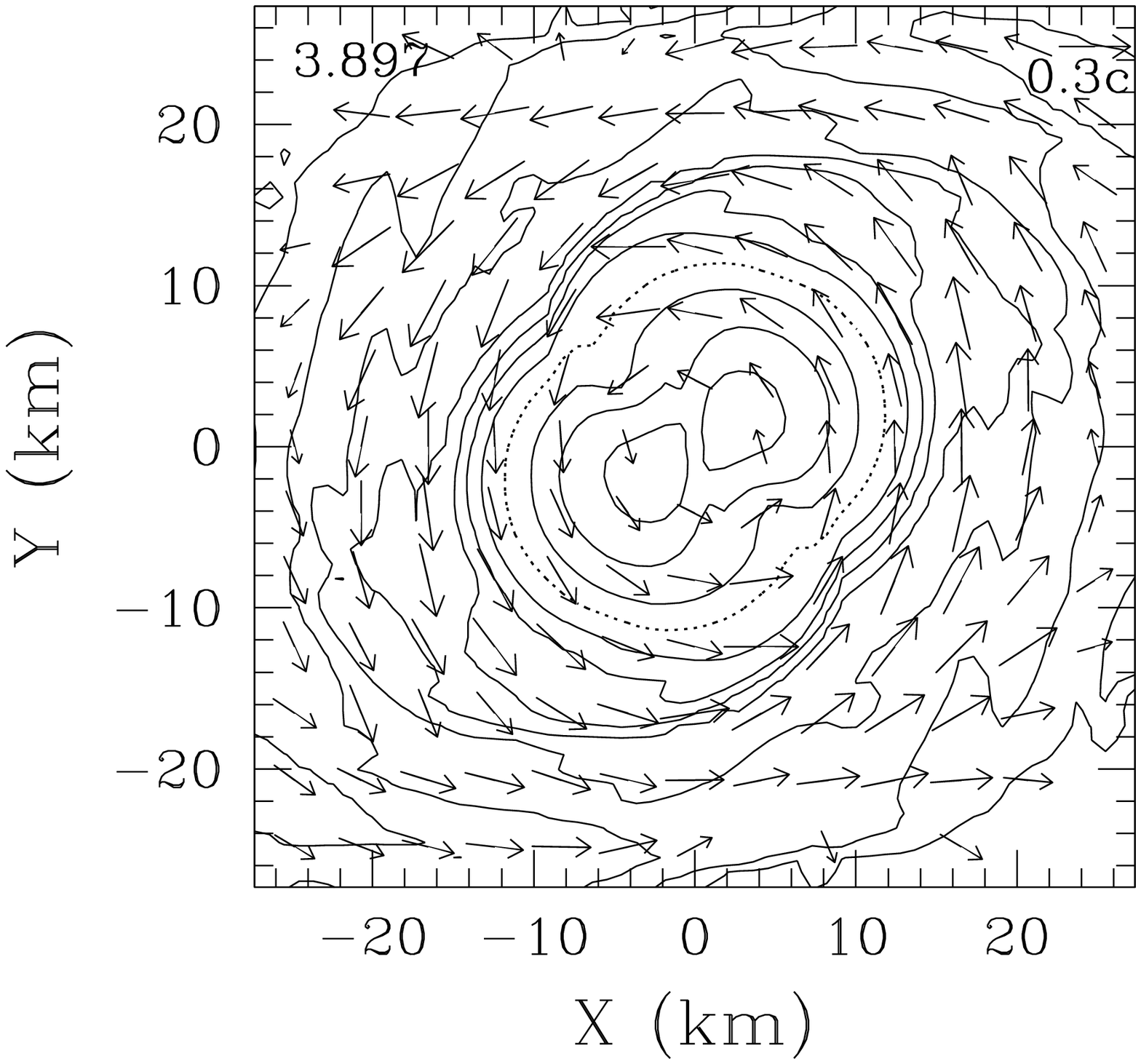}
   \includegraphics[height=5.0cm]{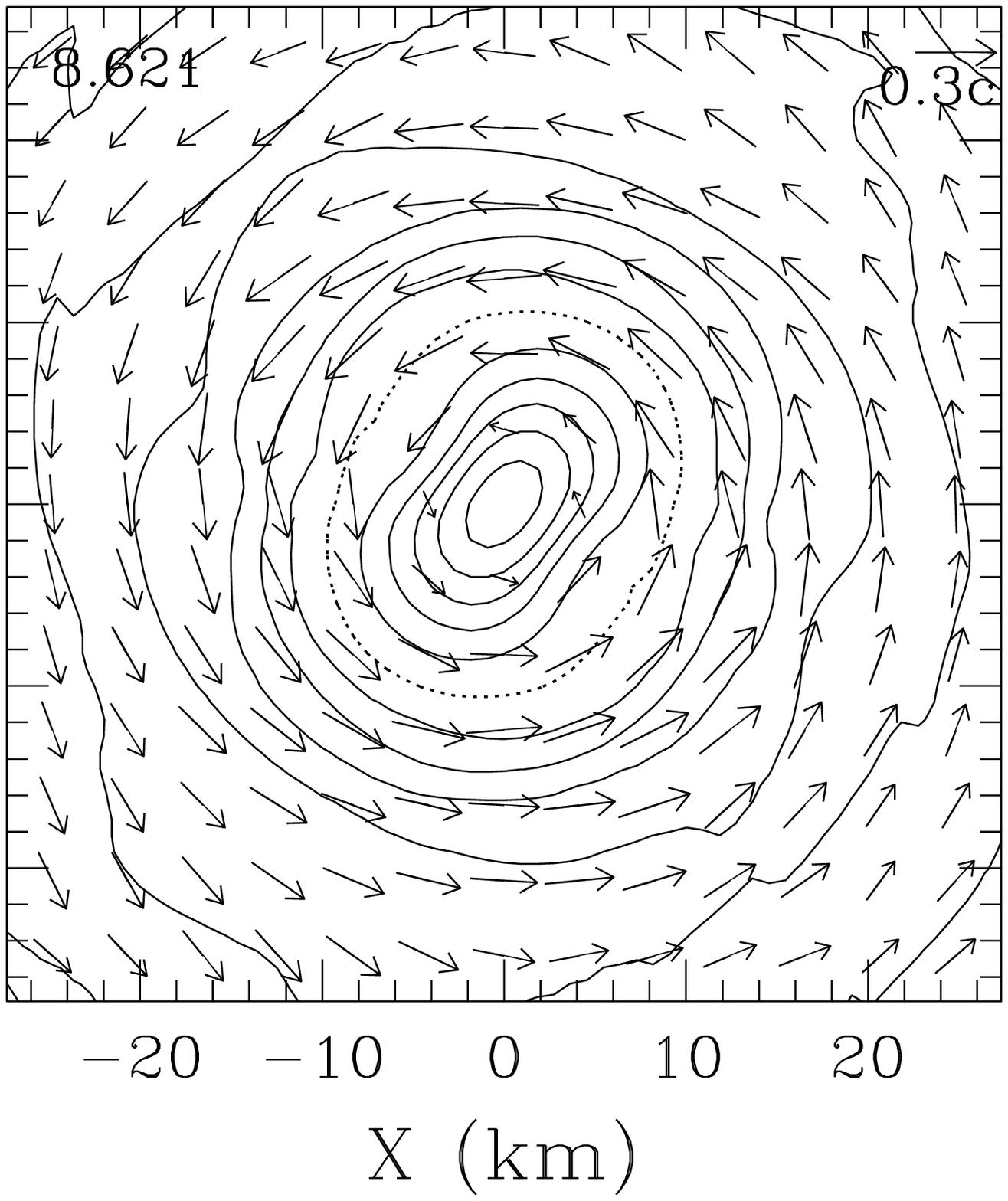}
   \caption{Formation of triaxial HMNS remnant following 
   NSNS merger in $2.7 M_{\odot}$ system. Snapshots of density contours 
   are shown in the equatorial plane.
   The number in the upper left-hand corner denotes the elapsed time 
   in ms; the initial orbital period is 2.11 ms. Vectors indicate 
   the local velocity field. 
   [From Shibata, Taniguchi and Ury{\= u} 2005.]}
   \label{fig2}
   \end{center}
   \end{figure}

The above scenario has become very relevant in light of the most
recent and detailed simulations in full general relativity of NSNS mergers.
State-of-the-art, fully relativistic simulations of NSNSs have been performed 
by Shibata and his collaborators 
(\cite[Shibata, Taniguchi \&  Ury{\= u} 2003,2005]{ShiTU03,ShiTU05}; 
\cite{Shi05}; \cite{ShiT06}). 
They consider mergers of $n=1$ polytropes,
as well as configurations obeying a more realistic nuclear 
equation of state (EOS). They treat mass ratios $Q_M$ in the 
range $0.9 \leq Q_M \leq 1$, consistent with the range of $Q_M$ in 
observed binary pulsars with accurately determined masses (\cite{ThoC99}; 
\cite{Sta04}). The key 
result is that there exists a critical mass 
$M_{\rm crit} \sim 2.5 - 2.7 M_{\odot}$ of the binary system above which the
merger leads to prompt collapse to a black hole, and below which the merger 
forms a hypermassive remnant. 
With the adopted EOS, the HMNS remnant undergoes delayed
collapse in about $\sim 100$ ms and emits a delayed gravitational wave burst.
Most interesting, prior to collapse, the remnant forms a triaxial 
bar when a realistic EOS is adopted (see Fig.~\ref{fig2}) and the bar 
emits quasiperiodic gravitational waves at a frequency $f \sim 3 - 4$ kHz.
Such a signal may be detectable by Advanced LIGO. 
It is interesting that for the adopted EOS, the mass $M_{\rm crit}$ is close
to the value of the total mass found in each of the observed binary pulsars.
Given that the
masses of the individual stars in a binary
can be determined by measuring the gravitational wave signal emitted 
during the adiabatic, inspiral epoch prior to plunge and merger,
the detection (or absence) of any
quasiperiodic emission from the hypermassive remnant prior to delayed 
collapse may significantly constrain models of the nuclear EOS. 

The possibility that a HMNS remnant forms 
following a NSNS merger had been foreshadowed in earlier Newtonian 
simulations (\cite[Rasio \& Shapiro 1994,1999]{RasS94,RasS99};
\cite{ZhuCM96}), in 
post-Newtonian simulations (\cite[Faber \& Rasio 2000,2002]{FabR00,FabR02}) 
and in conformally flat general relativistic simulations 
(\cite{FabGR04}). However, the recent fully  relativistic simulations by 
\cite[Shibata, Taniguchi \& Ury{\= u} (2003,2005)]{ShiTU03,ShiTU05}, 
\cite[Shibata (2005)]{Shi05} and \cite[Shibata \& Taniguchi (2006)]{ShiT06} 
provide the strongest theoretical evidence of this phenomenon to date, 
although the details undoubtedly 
depend on the adopted EOS.  Triaxial equilibria can arise only in 
stars that can support sufficiently high values of $\beta$  
exceeding the classical bifurcation point at $\beta \approx 0.14$; 
reaching such high values
requires  EOSs with adiabatic indicies exceeding $\Gamma \approx  2.25$ in 
Newtonian configurations, and
comparable values in relativistic stars. 
It is not yet known whether the true nuclear EOS in neutron stars 
is this stiff, or what agent for redistributing 
angular momentum in a hypermassive star dominates
(e.g., magnetic fields, turbulent viscosity or gravitational radiation).
But it is already evident that hypermassive stars are likely to form 
from some mergers and that they will survive many dynamical
timescales before undergoing delayed collapse. The recent 
measurement (\cite{NicSSLJKC05})
of the mass of a neutron star in a neutron star-white dwarf binary of 
$M= 2.1 M_{\odot}$ establishes an observational lower limit for 
the maximum mass of a neutron star; such a high value suggests that mergers in
typical NSNSs may form hypermassive stars more often than undergoing 
prompt collapse.

\subsubsection{Collapse of a Magnetized HMNS}

The HMNSs found above may survive for many rotation
periods.  However, as we stated, 
on longer timescales magnetic fields will transport
angular momentum and this will trigger gravitational collapse.
Two important magnetohydrodynamic (MHD) mechanisms which transport angular 
momentum are magnetic
braking (\cite{BauSS00,Sha00,CooSS03,LiuS04}) and the 
magnetorotational instability
(MRI; \cite{Vel59,Cha60}, \cite[Balbus \& Hawley 1991,1998]{BalH91,BalH98}). 
Magnetic breaking
transports angular momentum on the Alfv\'en time
scale, $\tau_A \sim R/v_A \sim 1 (B/10^{14}~{\rm
G})^{-1}~{\rm s}$,
where $R$ is the radius of the HMNS. MRI occurs wherever angular
velocity $\Omega$ decreases with cylindrical radius $\varpi$.  This
instability grows exponentially with an e-folding time of $\tau_{\rm
MRI} \approx 4 \left(\partial \Omega/\partial \ln \varpi
\right)^{-1}$, independent of the field strength. For 
typical HMNSs considered here, $\tau_{\rm MRI} \sim 1$~ms.  The length scale
of the fastest growing unstable MRI modes,
$\lambda_{\rm MRI}$, does depend on the field strength:
$\lambda_{\rm MRI}\sim 3~{\rm m}$ $(\Omega/4000 s^{-1})^{-1}$
$(B/10^{14}{\rm G}) \ll R$.
When the MRI saturates, turbulence consisting of small-scale eddies
often develops, leading to angular momentum transport on a
timescale much longer than $\tau_{\rm MRI}$. The computational
challenge of evolving an HMNS is having sufficient spatial grid to resolve
the MRI wavelength and sufficient integration time to follow the evolution
on the long Alfv\'en timescale.

To determine the final fate of the HMNS, it is necessary to carry out
MHD simulations in
full general relativity.  Such simulations have only recently become
possible.  \cite[Duez {\it et al.} (2005)]{DueLSS05} and 
\cite[Shibata \& Sekiguchi (2005)]{ShiS05} 
have  developed new codes to evolve the coupled set of 
Einstein-Maxwell-MHD equations self-consistently. Our two codes have since 
been used to simulate the evolution of magnetized HMNSs 
(\cite[Duez {\it et al.} 2006a,2006b]{DLSSS06a,DLSSS06b}), and
implications for short GRBs have been investigated (\cite{ShiDLSS06}). 
Both codes give very similar results.

We assume axisymmetry and equatorial symmetry in all of these simulations.
We use uniform computational grids with sizes
up to 500$\times$500 spatial zones in cylindrical coordinates.  To model the
remnant formed in binary merger simulations, we use as our initial data an
equilibrium HMNS constructed from a $\Gamma=2$ polytrope with mass
$M$ 1.7 times the spherical mass limit, or 1.5 times the limit for a uniformly
rotating star built out of the same EOS. These are the typical values expected
for a HMNS formed from a NSNS merger, for which $M \sim 2 \times 1.4 M_{\odot} 
= 2.8 M_{\odot}$.
The differential rotation profile is chosen so that
the ratio of equatorial to central $\Omega$ is $\sim 1/3$, comparable to what
is found in simulations of NSNS mergers.
(We find that an HMNS with a more realistic hybrid EOS rather than a 
polytrope evolves similarly; \cite{DueLSSS06b,ShiDLSS06})  We add a
seed poloidal magnetic field with strength
proportional to the gas pressure.  The initial magnetic pressure
is set much smaller than the gas pressure, but not so small that
$\lambda_{\rm MRI}$ cannot be resolved.  Therefore, we set
$\lambda_{\rm MRI} \approx R/10$, corresponding to $B \approx 10^{16}$ G
and max$(B^2/P) \sim 10^{-3}$. The resulting Alfv\'en timescale is
about $16$ central rotation periods, or $600M$.

\begin{figure}
\begin{center}
\includegraphics[width=1.7in]{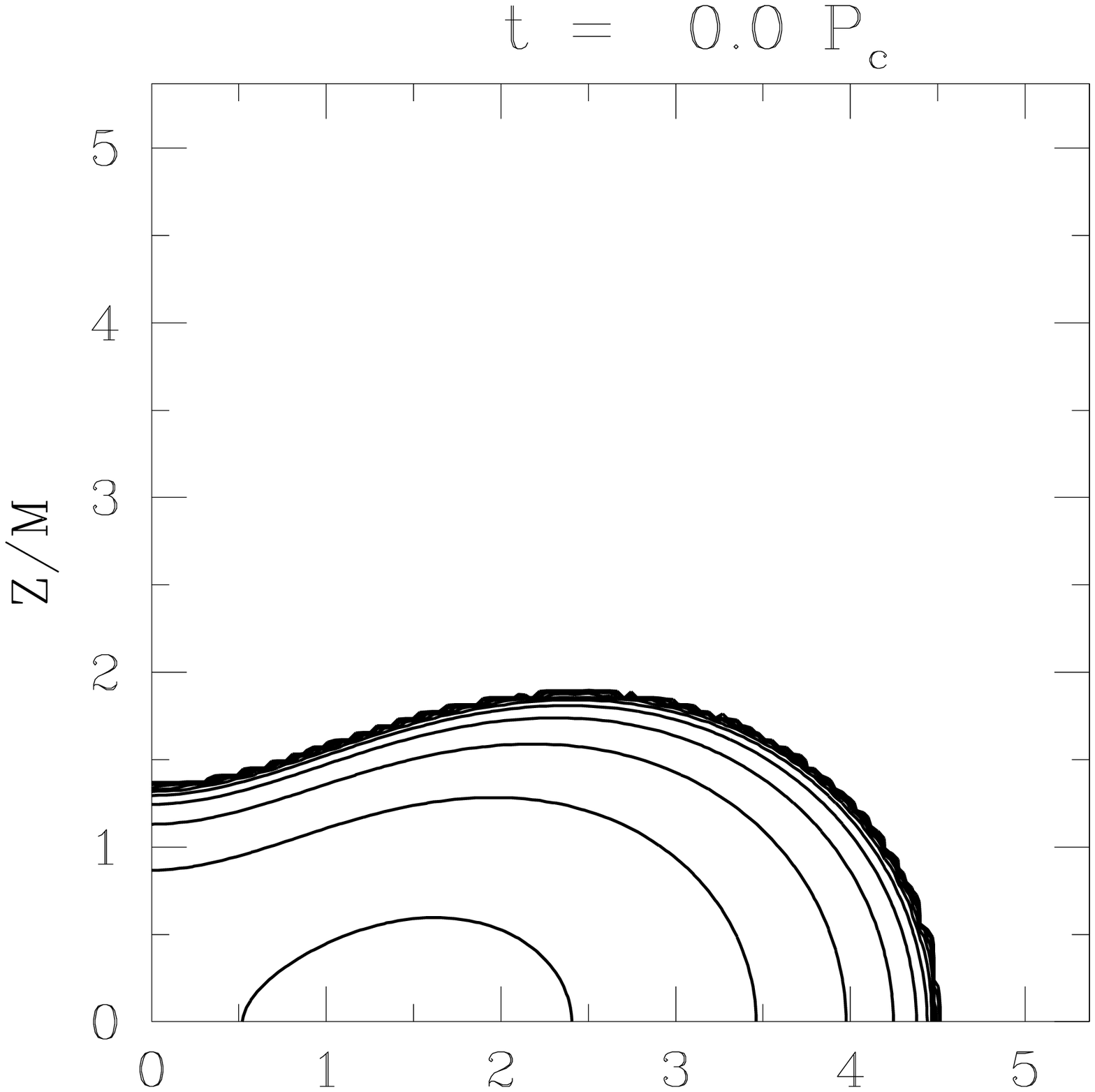}
\includegraphics[width=1.7in]{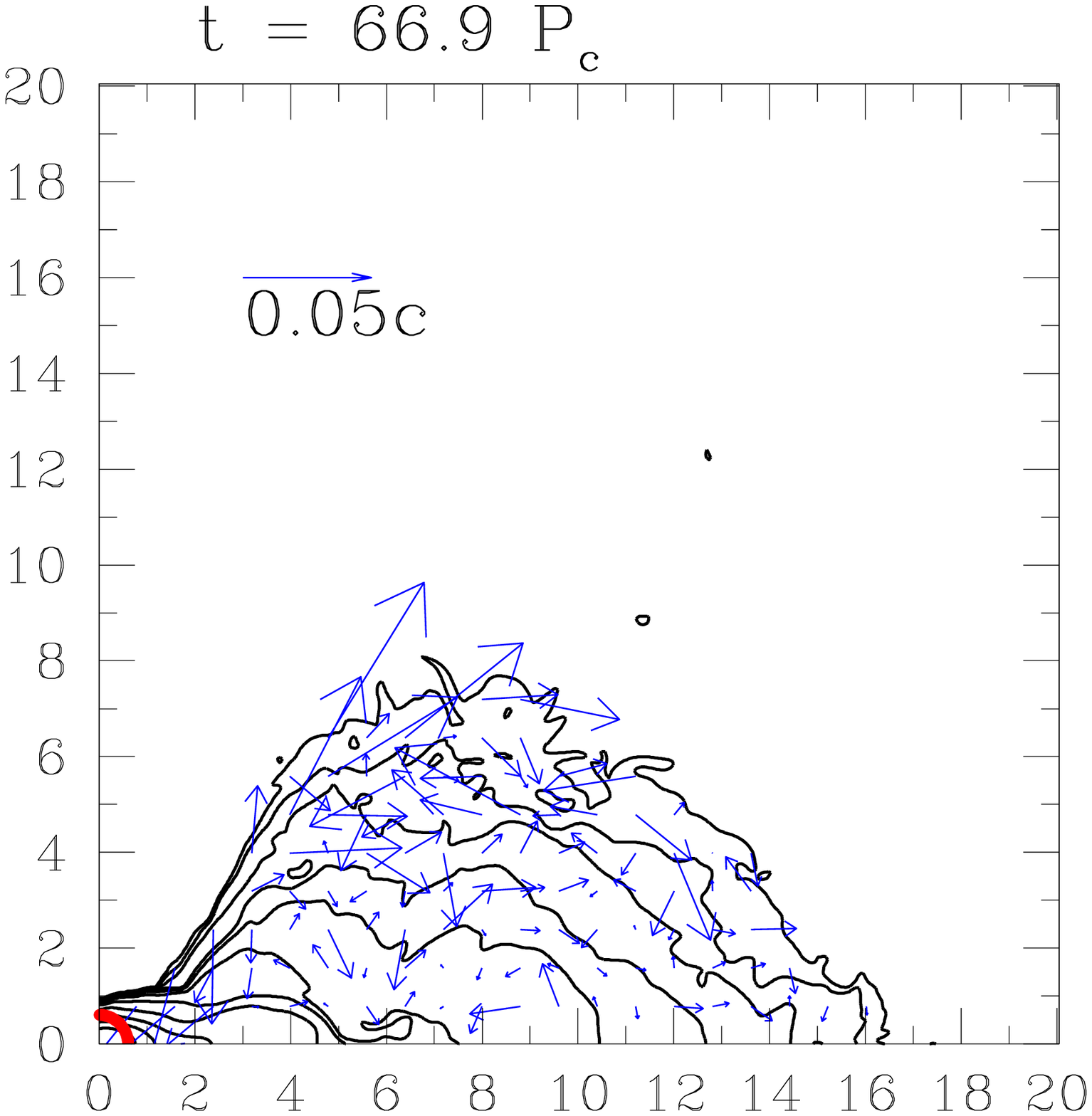}
\includegraphics[width=1.7in]{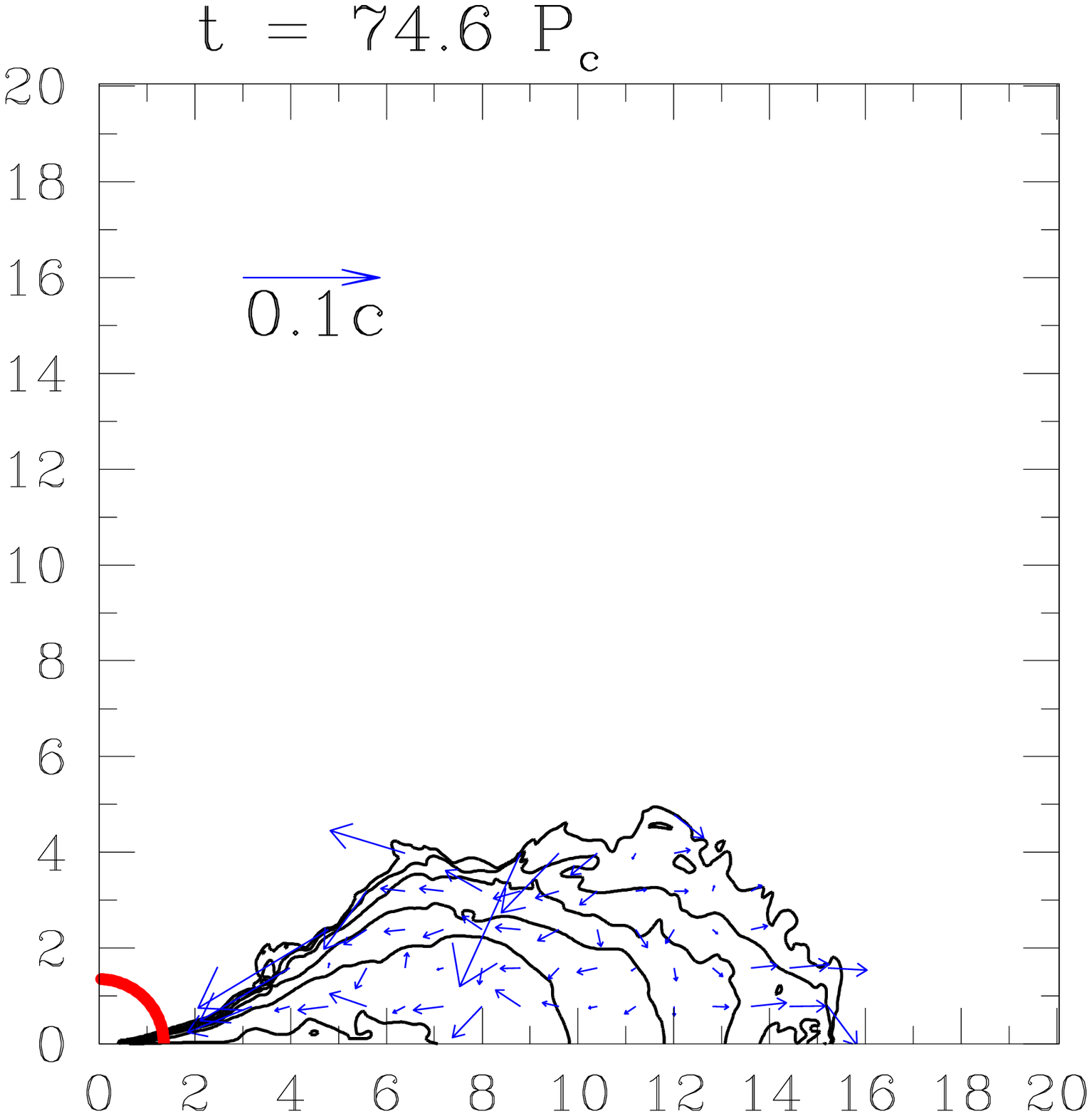} \\
\includegraphics[width=1.7in]{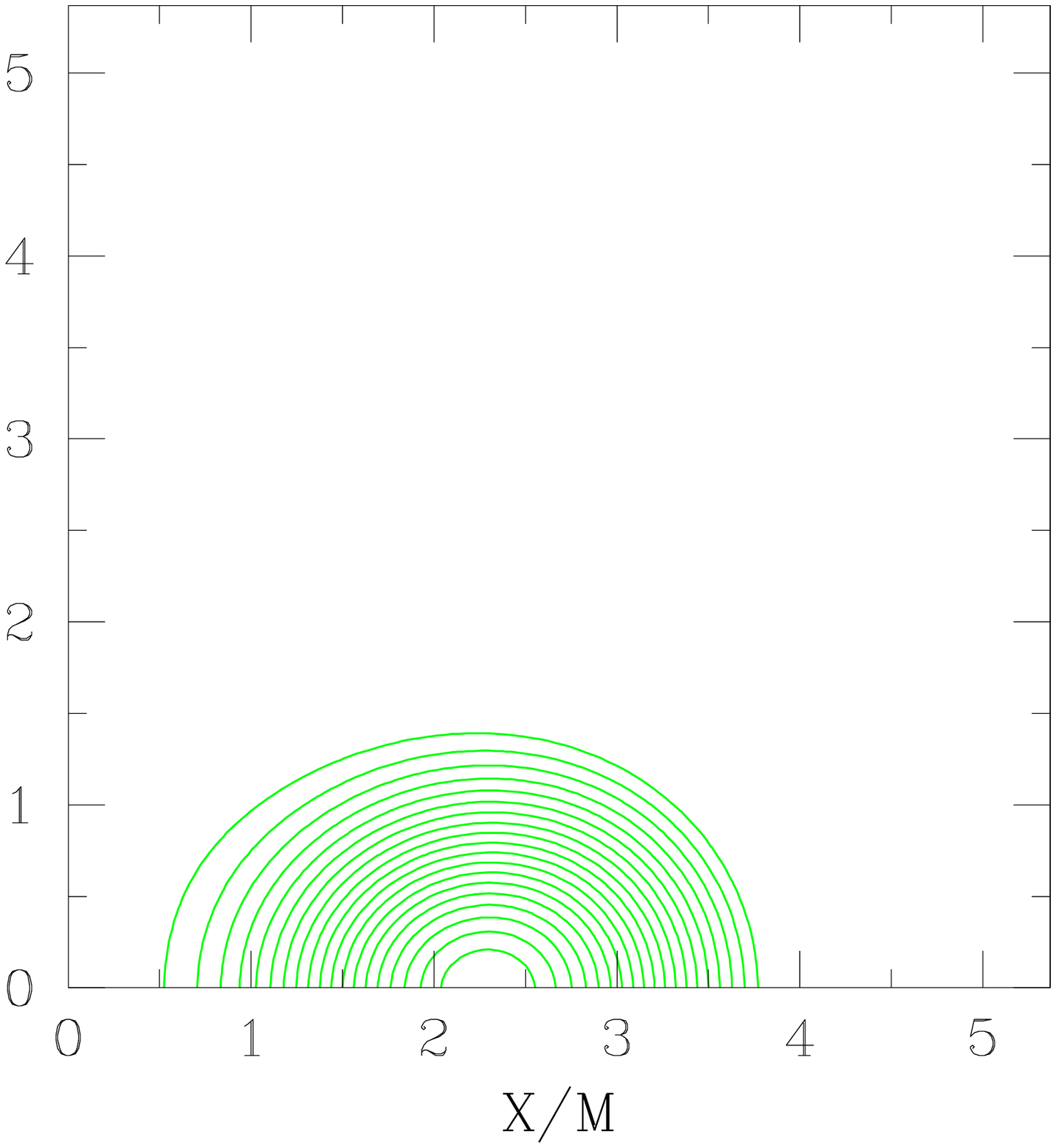}
\includegraphics[width=1.7in]{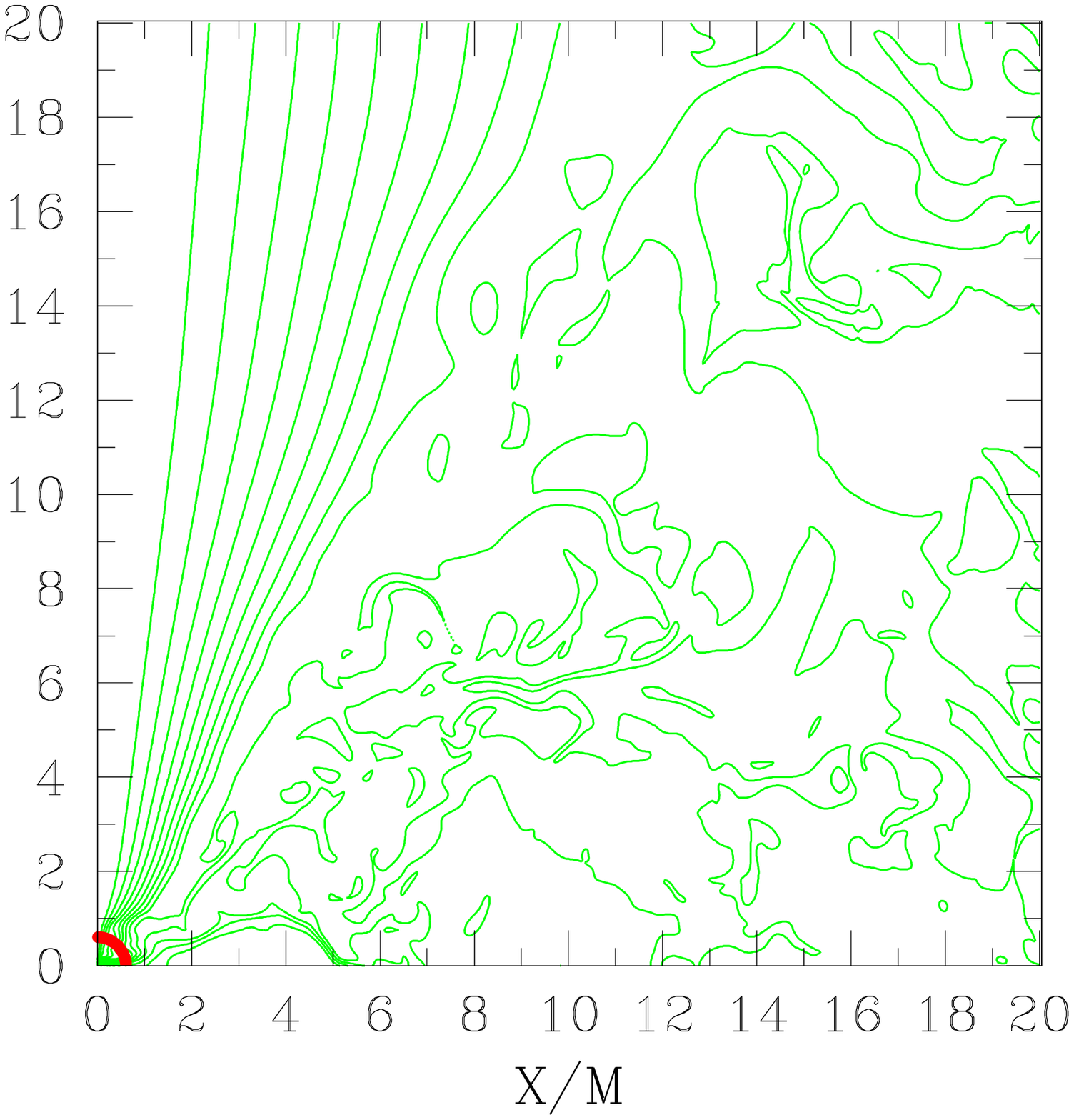}
\includegraphics[width=1.7in]{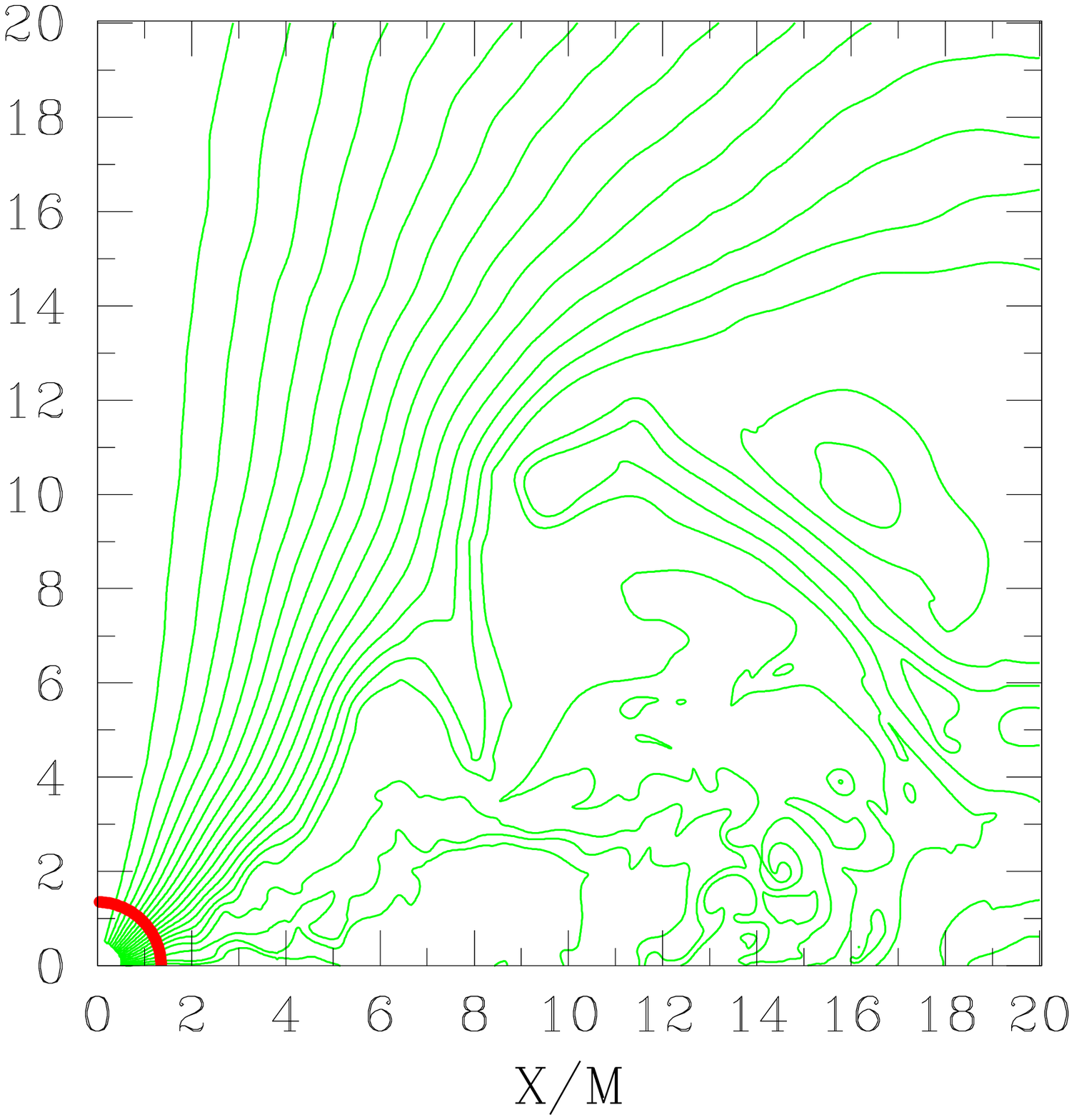} \\
\end{center}
\caption{Collapse of a magnetized HMNS to a black hole. 
The upper 3 panels show snapshots of the rest-mass density
contours and velocity vectors on the meridional plane. The lower
panels show the field lines (lines of constant vector potential $A_{\phi}$)
for the poloidal magnetic field at the same times
as the upper panels.
The density contours are drawn for $\rho/\rho_{\rm max,0}=
10^{-0.3 i-0.09}~(i=0$--12).
The field lines are drawn for $A_{\phi} = A_{\phi,\rm min}
+ (A_{\phi,\rm max} - A_{\phi,\rm min}) i/20~(i=1$--19),
where $A_{\phi,\rm max}$ and $A_{\phi,\rm min}$ are the maximum
and minimum value of $A_{\phi}$ respectively at the given time.
The thick solid curves in the lower left corner denote the apparent horizon.
[From \cite{DueLSSS06a}.]}
\label{duez}
\end{figure}

In our evolutions, the effects of magnetic winding are reflected in the
generation of a toroidal $B$ field which grows linearly with time
during the early phase of the evolution, and saturates on the Alfv\'en
timescale.  The effects of MRI are observed in an exponential growth 
of the poloidal field
on the $\lambda_{\rm MRI}$ scale, a growth which saturates after a few
rotation periods.  The magnetic fields cause angular momentum to be
transported outward, so that the core of the star contracts while the outer
layers expand.  After about 66 rotation periods, the core collapses to a
black hole.  Using the technique of black hole excision (see \cite{DueSY04} and 
references therein) to
remove the interior of the black hole, with its nasty spacetime singularity,
and replace it with suitable boundary conditions on all the variables just 
inside the horizon,
we continue the evolution to a quasi-stationary state.  
The final state consists of a
black hole of irreducible mass $\sim 0.9M$ surrounded by a hot accreting torus
with rest mass $\sim 0.1M$ and a magnetic field collimated along the 
rotation axis; see Fig.~\ref{duez}. 
At its final accretion rate, the torus should survive for $\sim$10ms.  The
torus is optically thick to neutrinos, and we estimate that it will emit
$\sim 10^{50}$ergs in neutrinos before being accreted.  We also find that
the cone outside the black hole centered along the rotation axis 
is very baryon-poor. All these properties make this system a promising central
engine for a short-hard GRB.

\section{Black Holes as Central Engines for GRBs}

\begin{figure}
\begin{center}
\includegraphics[scale=0.6]{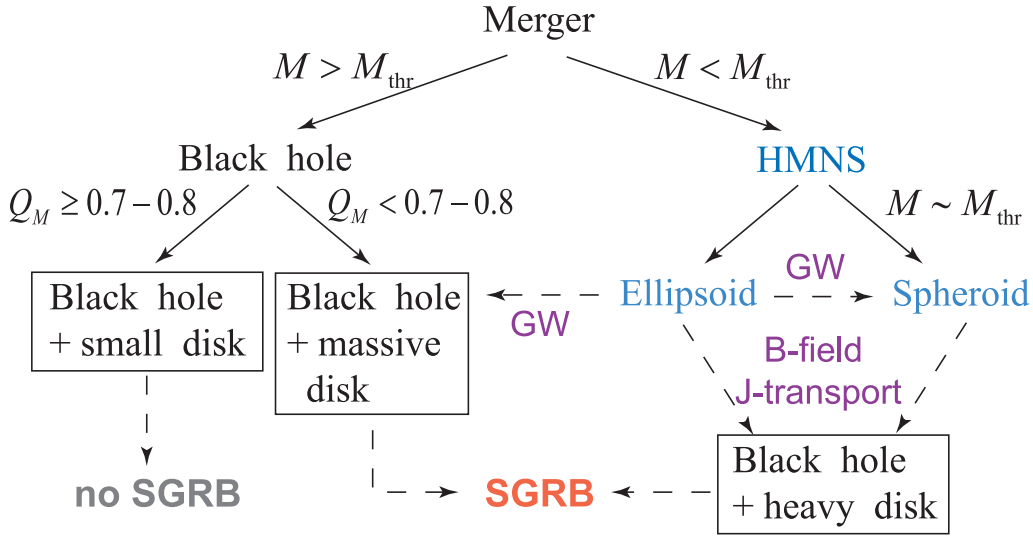}
\end{center}
\caption{Plausible routes for the formation of a short-hard gamma-ray
burst (SGRB) central engine following the merger of a binary neutron star.
Here ``GW'' (``B-field J-transport'') denotes angular momentum dissipation
dominated by gravitational wave emission (magnetic fields).
[From  \cite{ShiT06}.]}
\label{routes}
\end{figure}

The combined observations 
of {\it BATSE, Swift, HETE-2, Chandra} and the HST indicate
that GRBs comprise at least two classes: long-soft and short-hard. Long-soft
GRBs have characteristic timescales $\tau$ in the range $\tau \sim 2 - 10^3$s.
They are found in star-forming regions (spirals) and some are observed
to be associated with supernovae. The favored model for the progenitor of
a long-soft GRB is the collapse of a massive, rotating, magnetized 
star to a black hole ('collapsar'; \cite{MacW99}). By contrast, 
short-hard GRBs  
have characteristic timescales in the range $\tau \sim 10$ms $- 2$s.
They are identified in low star-forming regions (ellipticals) where
associations with supernovae can be excluded. The favored model for their
progenitors are either NSNS or BHNS mergers. 
Alternative routes by which NSNS mergers can result
in the generation of a short-hard GRB are traced in Fig.~\ref{routes}. These
alternatives have emerged from detailed simulations in general relativity.
The HMNS route has already been summarized in Section~\ref{bns}.

The inspiral and merger of NSNSs and BHNSs have important
implications for the detection of gravitational waves with Advanced LIGO.
Recent estimates for the rates for detectable NSNS mergers are promising, 
in the neighborhood of $20 - 30 {\rm yr}^{-1}$ (\cite{OshKKB06}). 
Simulations in general relativity now underway should be helpful in 
preparing for the exciting possibility of
the simultaneous detection of a gravitational wave {\it and} GRB from 
the {\it same} source in the near future.

\section{Magnetorotational Collapse of Massive Stars to Black Holes} 
\label{collapse}

Recently, we performed simulations in axisymmetry 
of the magnetorotational collapse
of very massive stars in full general relativity (\cite{LiuSS07}).
Our simulations are directly applicable to the collapse of
supermassive stars with masses $M \gtrsim 10^3M_{\odot}$ and to
very massive Population~III stars. They are also relevant 
for core collapse in massive Population I stars,
since in all of these cases the governing EOS up to the appearance of a 
black hole can be approximated by an adiabatic $\Gamma=4/3$ law (although
its physical origin is different).
These simulations may help explain the formation of 
the central engine
in the collapsar model of long-soft GRBs (\cite{MacW99}). 
Moreover, some long-soft GRBs observed at very high redshift might 
be related to the gravitational collapse
of very massive Pop~III stars (\cite{SchGF02,BroL06}).
Hence these simulations may also provide
direct insights into the formation of GRB central engines arising 
from first-generation stars.

\begin{figure}
\begin{center}
\includegraphics[width=2.0in]{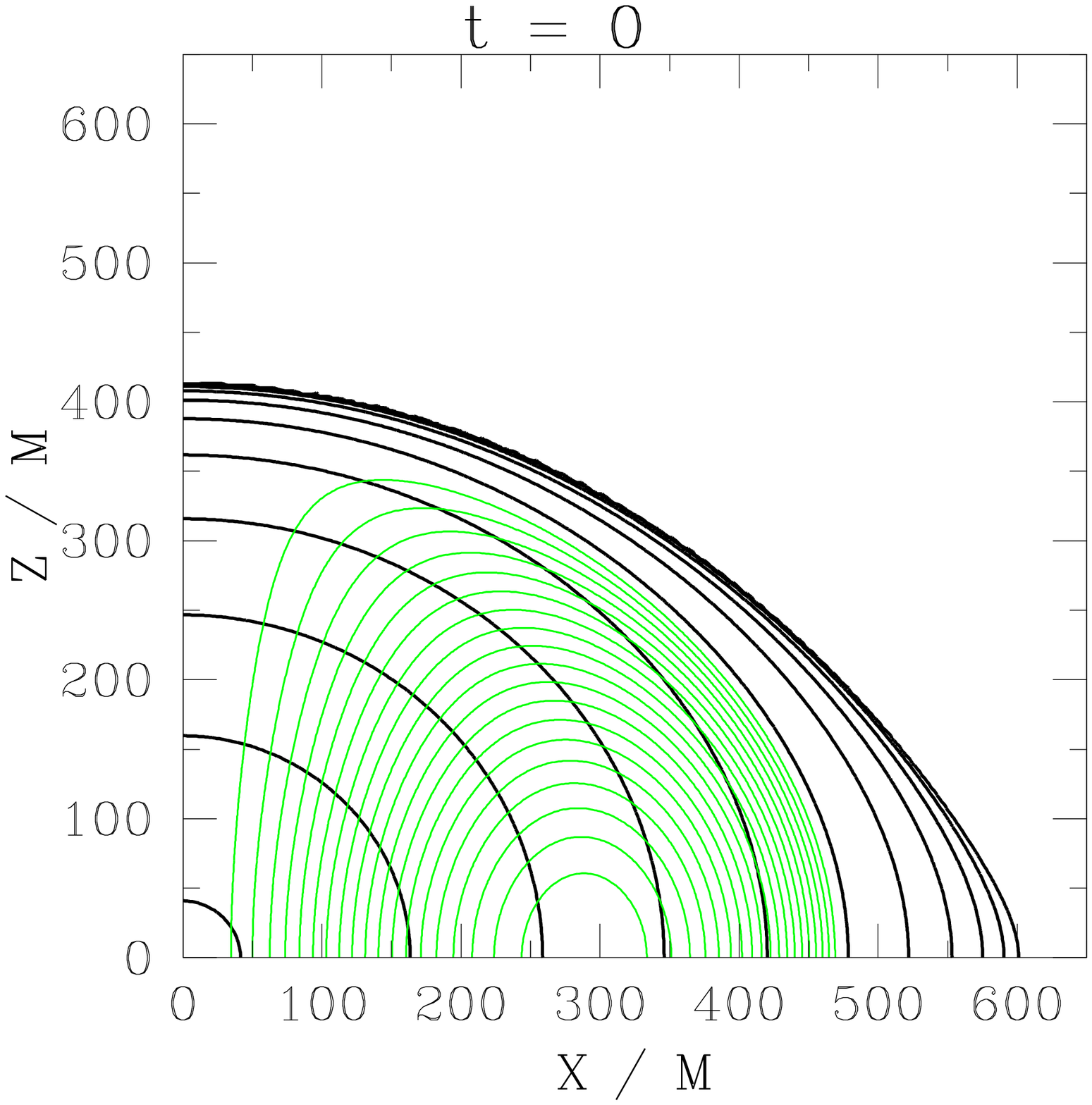}
\includegraphics[width=2.0in]{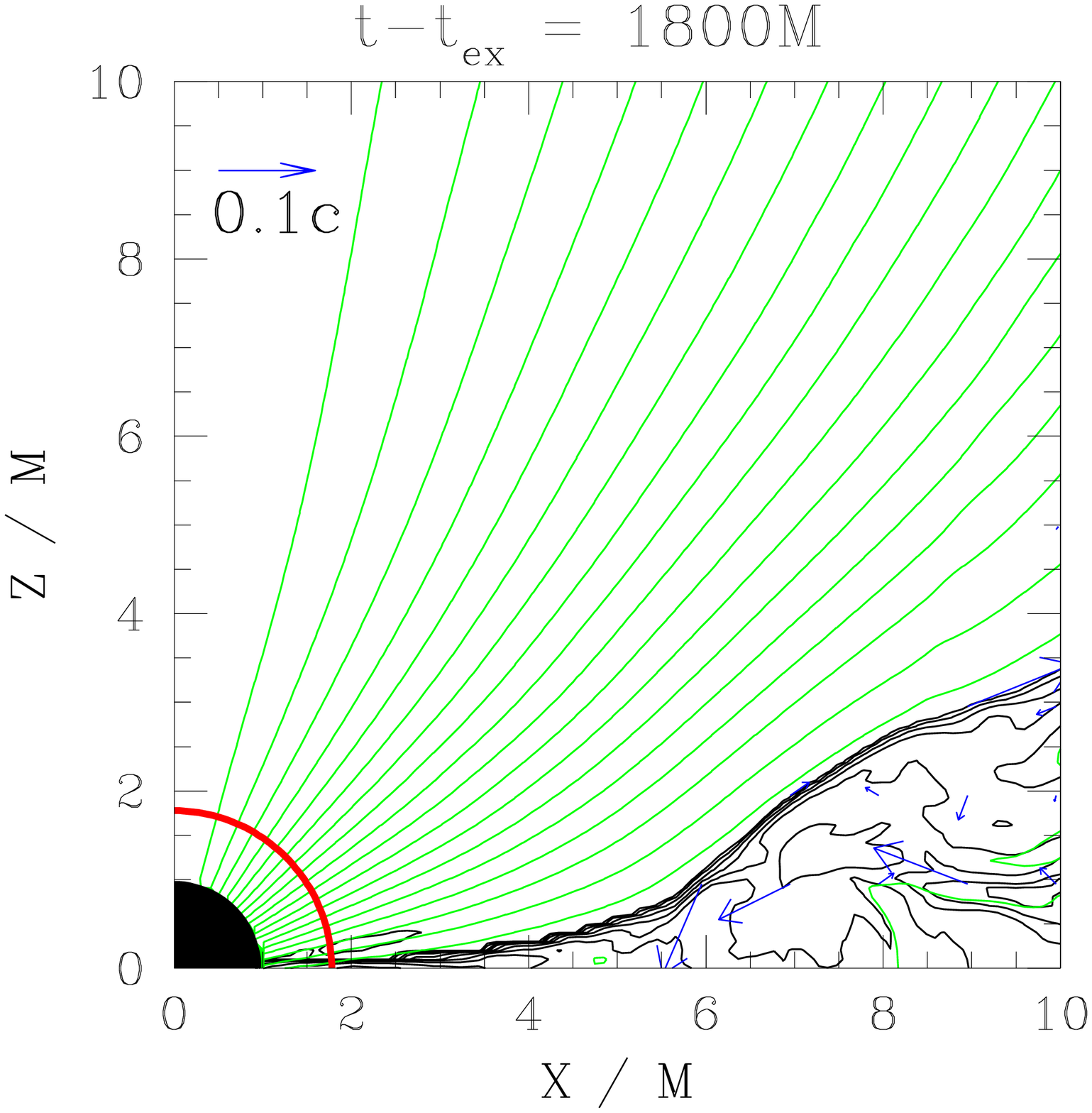}
\end{center}
\caption{
Magnetoroational collapse of a massive star to a black hole. 
Snapshots of meridional rest-mass density contours, velocity vectors
and poloidal magnetic field lines for the
initial and endpoint configurations for $n=3$ collapse.
Field lines coincide with contours of vector potential $A_{\varphi}$ and 
are drawn for
 $A_{\varphi}=A_{\varphi,{\rm max}}(j/20)$ with $j=1, 2,
\cdots, 19$ where $A_{\varphi,{\rm max}}$ is the maximum value 
of $A_{\varphi}$. In the final, post-excision model ($t_{\rm ex} = 29150M$),
the density levels are drawn
for $\rho_0 = 100 \rho_c(0) 10^{-0.3j}~(j=0$--10). The thick
arc near the lower left corner of the right-hand frame 
denotes the apparent horizon and the shaded
region the excision domain.
[Adapted from \cite{LiuSS07}.]}
\label{LiuSS07} 
\end{figure}

The simulations of \cite[Liu, Shapiro \& Stephens (2007)]{LiuSS07} model 
the initial configurations by $n=3$ polytropes, uniformly rotating near the
mass-shedding limit and at the onset of radial instability to collapse.
These simulations extend the earlier results of
\cite[Shibata \& Shapiro 2002]{ShiS02} by incorporating the 
effects of a magnetic field and by
tracking the evolution for a much longer time after the appearance
of a central black hole.
The ratio of magnetic to rotational kinetic energy in the initial 
stars is chosen to be small (1\% and 10\%). We find that such
magnetic fields do not affect the {\it initial} collapse significantly.
The core collapses to a black hole,
after which black hole excision is employed to continue the evolution
long enough for the hole reach a quasi-stationary state.
We find that the black hole mass is $M_h = 0.95M$ and its spin parameter is
$J_h/M_h^2 = 0.7$, with the remaining matter forming a torus
around the black hole.  The {\it subsequent} evolution of the torus does 
depend on the strength of the magnetic field. We freeze the spacetime metric
(``Cowling approximation'') and continue to follow the evolution of the
torus after the black hole has relaxed to quasi-stationary equilibrium.
In the absence of magnetic fields,
the torus settles down following ejection of a small amount of matter due
to shock heating. When magnetic fields are present,
the field lines gradually collimate along the
hole's rotation axis. MHD shocks and the
MRI generate MHD turbulence in the
the torus and stochastic accretion onto the central
black hole (see Fig.~\ref{LiuSS07}). When the magnetic field is 
strong, a wind is generated in
the torus, and the torus undergoes radial oscillations that drive
episodic accretion onto the hole.
These oscillations produce long-wavelength
gravitational waves potentially detectable by 
LISA.
The final state of magnetorotational collapse always
consists of a central black hole
surrounded by a collimated magnetic field and a hot, thick accretion
torus. This system is a viable candidate for the
central engine of a long-soft gamma-ray burst.

\section{Cosmological Growth of Supermassive Black Holes}

Growing evidence indicates that supermassive black holes (SMBHs) 
with masses in the range  $ 10^6 - 10^{10} M_{\odot}$ exist and 
are the engines that power AGNs and quasars.
There is also ample evidence that SMBHs reside at the centers
of many, and perhaps most, galaxies,
including the Milky Way.
The highest redshift of a quasar discovered to 
date is $Z_{\rm QSO} = 6.43$, corresponding to
QSO SDSS 1148+5251 (\cite{Fan03}). Accordingly, if they are the 
energy sources in quasars (QSOs), the first SMBHs
must have formed prior to $Z_{\rm QSO} = 6.43$, or within $t = 0.87$ Gyr 
after the Big Bang in the concordance $\Lambda$CDM cosmological model. 
This requirement sets a significant constraint on black
hole seed formation and growth mechanisms in the early universe.
Once formed, black holes grow by a combination of
mergers and gas accretion.

The more massive the initial seed, the less time is required for it 
to grow to SMBH scale and the easier it is to have a SMBH in 
place by $Z \geq 6.43$. One possible progenitor
that readily produces a SMBH is a supermassive
star (SMS) with $M \gg 10^3 M_{\odot}$ (\cite{Ree84,Sha04}). 
SMSs can form when gaseous 
structures build up sufficient radiation pressure to inhibit 
fragmentation and prevent normal star formation; plausible cosmological 
scenarios have been proposed that can lead to
this situation (\cite{Gne01,BroL03}).
Alternatively, the seed black holes that later grow to become
SMBHs may originate from the collapse of Pop III 
stars $\lesssim 10^3 M_{\odot}$ (\cite{MadR01}). 
To achieve the required growth to $\sim 10^9 M_{\odot}$ 
by $Z_{\rm QSO} \gtrsim 6.43$, it may be
necessary for gas accretion, if restricted by the Eddington limiting
luminosity, to occur 
at relatively {\it low} efficiency of rest-mass to radiation energy conversion 
($\lesssim 0.2$; \cite{Sha05} and references therein), as we discuss
below.

The efficiency of black hole accretion, and the resulting rate of black hole
growth, is significantly affected by the spin of the black hole.
The spin evolution of a black hole begins at birth. If the hole arises
from the collapse of a massive or supermassive star, then it is likely
to be born with a spin parmeter in the range $0 \leq a/M \lesssim 0.8$.
The lower limit applies if the progenitor star (or core) is nonrotating, 
the upper limit if it is spinning uniformly at the mass-shedding 
limit at the onset of collapse, as found in the simulations of
\cite[Shibata \& Shapiro]{ShiS02} and \cite[Liu \& Shapiro 2007]{LiuS07} and
discussed in Section~\ref{collapse}. Major mergers with other black holes
of comparable mass will cause the black hole to spin up suddenly to
$a/m \approx 0.8 - 0.9$, as the merged remnant acquires almost all of the  
mass and angular momentum that characterizes a
circular orbit, quasistationary BHBH binary at the innermost stable circular
orbit (ISCO). Once secular gravitational radiation loss 
drives the binary past the ISCO, the black holes 
undergo a rapid dynamical plunge and coalescence, with little additional 
loss of energy and angular momentum..
This anticipated behavior has now been confirmed by numerical 
simulations of BHBH mergers (see Centrella 2007, This Volume,
and references therein). 
These simulations also show that gravitational
radiation reaction can induce a large kick velocity ($\gtrsim 1000$~km/s)
in the remnants following mergers.
While in principle these large kick velocities  pose a great hazard for the 
growth of
black hole seeds to SMBHs by $Z \sim 6$, large
kicks are possible only if the spins of
the black hole binary companions are appreciable and their
masses are comparable. In the end, gravitational recoil does not pose a
significant threat to the formation of the SMBH population observed locally, 
although high mass seeds are favored (\cite{Vol07}). 

Minor mergers with many
smaller black holes, isotropically distributed, cause the black hole
to spin down: $a/m \sim M^{-7/3}$ (\cite{HugB03,GamSM04}). 
The reason is that the ISCO and
specific angular momentum of black holes orbiting counter-clockwise 
is larger than for holes orbiting clockwise, hence the net effect of
isotropic capture is spindown.

However, it is likely that 
most of the mass of a supermassive black hole has been 
acquired by gas accretion, not mergers. Such a conclusion 
can be inferred from the observation that the luminosity density
of quasars is roughly $0.1-0.2$ of the local SMBH mass density 
(\cite{Sol82,YuT02}), 
an equality that arises naturally from growth via gaseous disk accretion. 
Steady gas accretion will quickly drive the black hole to spin equilibrium,
with a spin parameter that depends on the nature of the flow. 
If accretion occurs via a
relativistic ``standard thin disk", then the hole will be driven to the 
Kerr limiting value, $a/M =1$ (\cite{Bar70}). Correcting for the recapture
of some of the emitted photons from such a disk reduces the equilibrium spin 
value to $0.998$ (\cite{Tho74}). If, however, the accretion is driven 
by MRI turbulence in a relativistic MHD disk, then recent simulations indicate (\cite{McKG04,GamSM04,DeVHKH05}) that the equilibrium spin will fall 
to $\sim 0.95$. The small differences between these equilibrium spin 
parameters is deceptive, for they correspond to very different 
rest-mass-to-radiation conversion efficiencies,
as shown in Table~\ref{spin}.

\begin{table}
  \begin{center}
  \begin{tabular}{lccc}
$a/M$  & $\epsilon_M$ & Spin Equilibrium? & Characterization\\[3pt]
 0.0   & 0.057        &  no               & standard thin disk; nonspinning BH\\
 0.95  & 0.19         &  yes              & turbulent MHD disk\\
 0.998 & 0.32         &  yes              & standard thin disk; photon recapture\\
 1.0   & 0.42         &  yes              & standard thin disk; max spin BH\\
  \end{tabular}
  \caption{Rest-mass-to-radiation conversion efficiency {\it vs.} black hole spin.}
  \label{spin}
  \end{center}
\end{table}

The significance of these different values is that the growth of
a black hole with time via steady accretion depends 
{\it exponentially}, on the rest-mass-to-radiation conversion efficiency,
$\epsilon_M$, as we will now recall. 
Define the rest-mass-to-radiation conversion efficiency 
$\epsilon_M$ and the luminosity efficiency $\epsilon_L$ according to
\begin{equation}
\epsilon_M \equiv L/\dot M_0 c^2 = \epsilon_M(a/M)
~~~~\epsilon_L \equiv L/L_E\,
\end{equation}
where $M$ is the black hole mass, $M_0$ is the accreted rest-mass, and 
$L_E$ is the Eddington luminosity given by
\begin{equation}
L_E =\frac{4 \pi M \mu_e m_p c}{\sigma_T}
\approx 1.3 \times 10^{46} \mu_e M_8 ~{\rm erg~s^{-1}}\ . 
\end{equation}
Here $\tau$ is the growth timescale,
\begin{equation}
\tau = \frac{M c^2}{L_E} \approx 0.45 \mu_e^{-1} ~{\rm Gyr}\ ,
\end{equation}
and $\mu_e$ is the mean molecular weight per electron.
With the above definitions the growth rate
of a black hole due to accretion is
\begin{equation} \label{dmdt}
\frac{dM}{dt} = (1-\epsilon_M) \frac{dM_0}{dt} =
 \left[\frac{\epsilon_L (1-\epsilon_M)}{\epsilon_M} \right]
\frac{M}{\tau}\ .
\end{equation}
Integrating equation~(\ref{dmdt}) for steady accretion with constant 
efficiencies trivially yields the mass amplification of a black hole with time,
\begin{equation} \label{amp}
M(t)/M(t_i)= {\rm exp} \left[\frac{\epsilon_L (1-\epsilon_M)}{\epsilon_M}
\frac{(t-t_i)}{\tau} \right]\ ,
\end{equation}
showing the exponential dependence on the efficiency 
factors. 

Now a possible clue to the upper limit of $\epsilon_L$
is provided by the broad-line quasars in
a Sloan Digital Sky Survey sample of 12,698 nearby 
quasars in the redshift
interval $0.1 \leq z \leq 2.1$. 
This survey supports the value $\epsilon_L \approx 1$
as a physical upper limit (\cite{McLD04}).
Hence to maximize mass amplification via
steady accretion, one must accrete with a small value of $\epsilon_M$
at the Eddington limit, $\epsilon_L \approx 1$.

\begin{figure}
\begin{center}
\includegraphics[scale=0.5]{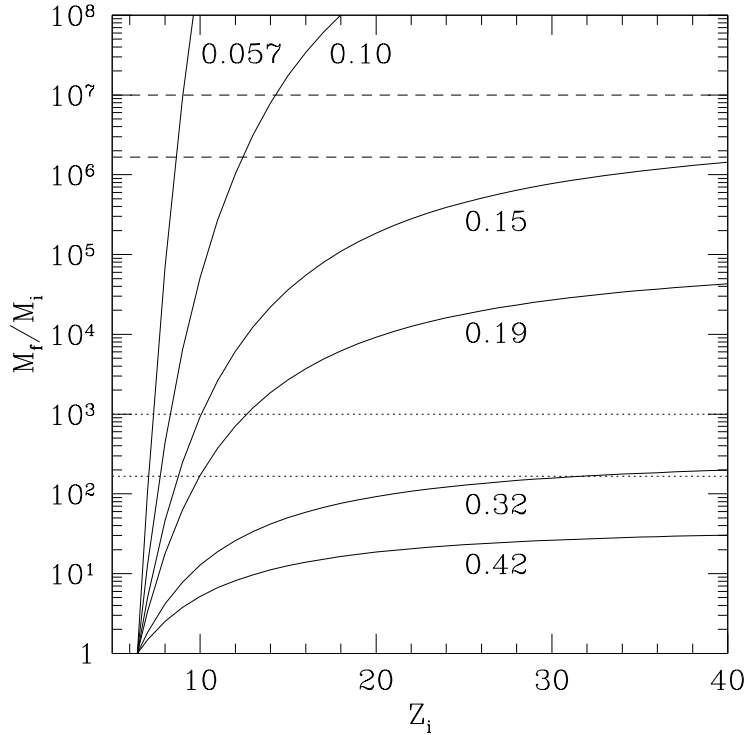}
\end{center}
\caption{Black hole accretion mass amplification $M_f/M_i$ versus redshift $Z_i$
of the initial seed. Here we plot the amplification achieved by
redshift $Z_f = 6.43$, the highest known quasar redshift, corresponding to
1148+5251. Each solid curve is labeled by the adopted
constant radiation efficiency, $\epsilon_M$; the luminosity is assumed to be the Eddington
value ($\epsilon_L =1$). The horizontal dashed lines bracket the
range of amplification required for accretion alone to grow a
seed black hole of mass
$100 \le M/M_{\odot} \le 600$ formed from the collapse of a Pop III star to $10^9 M_{\odot}$.
The horizontal dotted lines bracket the required accretion amplification range
assuming that mergers account for a growth of $10^4$ in black hole mass, the remainder being by gas accretion.
[From  \cite{Sha05}.]}
\label{accrete}
\end{figure}

Fig.~\ref{accrete} evaluates equation~(\ref{amp}) for black hole seeds that 
form at redshift $Z_i$ from the collapse of
Pop III stars in the range $100 \lesssim M/M_{\odot} \lesssim 600$, and
subsequently grow to $10^9 M_{\odot}$ by $Z_{\rm QSO} = 6.43$. The 
$\Lambda$CDM concordance cosmological model is assumed.
Of particular
relevance is the case of ``merger-assisted'' mass amplification, whereby 
mergers account for a typical growth
of $\sim 10^4$ in black hole mass, the remainder being by
gas accretion (see, e.g., \cite{YooM04}). 
The figure shows that for $Z_i \gtrsim 40$, the required growth of the seed to
SMBH status is easily achieved for
a relativistic MHD accretion disk, but is only marginally 
possible for a standard thin
disk that accounts for photon recapture,
and not at all possible for a standard
thin disk that drives the black hole to maximal spin.
However, if the initial black hole seed is less than $600 M_{\odot}$, 
accretion via a standard thin
disk appears to be ruled out altogether.
The fact that a black hole driven to spin equilibrium
by a turbulent MHD disk accretes with low enough efficiency to grow
to supermassive size by $Z_{\rm QSO} \approx 6.43$ is potentially
significant. It points to the need
for further relativistic MHD simulations of black hole accretion 
with ever greater physical sophistication, including the full effects of 
radiation transfer.

These same conclusions also hold
if the black hole seed forms much earlier
than $Z_i \approx 40$, and they may be tightened if the seed forms later.
In fact, it may be likely that the seed forms later, at $Z_i \lesssim 40$, given
that even $4-\sigma$ peaks in the density perturbation spectrum
for the progenitor halo of SDSS 1148+5251
do not collapse until $Z \sim 30$ in the $\Lambda$CDM concordance cosmology
(see, e.g. Figure 5 in \cite{BarL01}.) Moreover, the potential wells of
the earliest halos are quite shallow ($\sim 1$ km/s) and may not be 
able to retain
enough gas to form stars. Nevertheless, the effect of altering the date of birth
of the black hole seed is not very great unless $Z_i \lesssim 20 - 25$, as is evident from Fig. 3.

Should a quasar be discoverd 
at $Z_{\rm QSO}$ substantially above $6.43$,
it would not be understood easily in the context of
supermassive black hole growth by gas
accretion from a seed arising from the collapse of a 
stellar-mass, Pop III progenitor. 

The analysis presented here is illustrative only; the results may
change as the treatment is refined. 
The main point of this example is to emphasize that our understanding 
of structure formation in the early universe as it pertains to 
the formation and growth of supermassive black holes 
(global physics) depends in part on resolving some of the 
important details of relativistic BHBH recoil and 
relativistic black hole accretion (local physics). To understand
these details, in turn, requires  
large-scale simulations in full general relativity, which are now
possible and underway.

\section{Conclusions}

We have discussed a number of black hole scenarios
that require numerical simulations in full general relativity for
true understanding. Some of these processes involve the formation
of black holes, others concern their subsequent growth and 
interactions. The
important point is that numerical relativity is at last mature
enough to probe these issues reliably. Numerical relativity can now serve as 
an important tool to simulate 
stellar collapse to black holes,
the inspiral, merger and recoil of BHBHs, NSNSs and BHNSs,
the generation of gravitational waves from stellar collapse and 
binary mergers, accretion onto black holes,
and countless other phenomena involving black holes and their strong
gravitational fields. With such a tool,
these processes now can be investigated at a fundamental 
level without many of
the ad hoc assumptions and approximations 
required in previous treatments. This should lead to an improved
qualitative picture, as well as more reliable quantitative
results.

We are grateful to T. Baumgarte, C. Gammie, Y-T Liu, and M. Shibata for useful
discussions.  S.L.S is supported by NSF grants PHY-0205155, 
PHY-0345151, and PHY-0650377 
and NASA Grants NNG04GK54G and NNX07AG96G to the University of Illinois.

\end{document}